\documentclass{PoS}
\usepackage{epstopdf}
\title{High-luminosity LHC prospects with the upgraded ATLAS detector}

\ShortTitle{High-luminosity LHC prospects with the upgraded ATLAS detector}

\author{\speaker{Magdalena Slawinska}%
        \thanks{On behalf of the ATLAS Collaboration.}\\
       Nikhef National institute for subatomic physics\\
       E-mail: \email{Magdalena.Slawinska@nikhef.nl}}

\abstract{Run 1 at the LHC was very successful with the discovery of a new boson. The boson's properties are found to be compatible with those of the Standard Model Higgs boson. It is now revealing the mechanism of electroweak symmetry breaking and (possibly) the discovery of physics beyond the Standard Model that are the primary goals  of the just restarted LHC. The ultimate precision will be reached at the High-Luminosity LHC phase  with a proton-proton  centre-of-mass energy of 14 TeV. In this contribution physics prospects are presented for ATLAS for the integrated luminosities 300 and 3000 fb$^{-1}$: the ultimate precision attainable on measurements of the Higgs boson couplings to elementary fermions and bosons, its trilinear self-coupling, as well as perspectives on the searches for supersymmetric partners associated with the Higgs boson. Benchmark studies are presented to show how the sensitivity improves at the future LHC runs. For all these studies, a parameterised simulation of the upgraded ATLAS detector is used and expected pileup conditions are accounted for.}

\FullConference{XXIV International Workshop on Deep-Inelastic Scattering and Related Subjects\\
         11-15 April, 2016\\
         DESY Hamburg, Germany}

\begin{document}
%\linenumbers
%%%%%%%%%%%%%%%%%%%%%%%
\section{Introduction}

During Run 1 of the LHC a new boson was discovered by the ATLAS and CMS collaborations~\cite{Aad:2012tfa, Chatrchyan:2012xdj}.
Its measured properties are largely compatible with those of the SM Higgs boson~\cite{ATL+CMS} and precise measurements are
instrumental in determining if it is indeed  "the" Higgs predicted by~\cite{Higgs mechanism} and 
responsible for electroweak symmetry breaking.

After a very successful Run 1 of the LHC,  Run 2 data taking is currently taking place with
proton-proton collisions at a centre-of-mass energy $\sqrt{s}=$ 13 TeV. ATLAS is expected to collect  
an integrated luminosity of about 100~fb$^{-1}$ by the end of 2018. 
Starting in  2021, proton-proton collisions at $\sqrt{s}=14$~TeV are expected during the Run 3 of the LHC
amounting to about 300~fb$^{-1}$.
In 2016  the High Luminosity (HL-LHC) phase will start.
With an instantenuous luminosity  of $7.5 \times 10^{34}$~cm$^{-2}$s$^{-1}$ 
about 3000~fb$^{-1}$  are expected to be collected 
over a period of about ten years. 
Collisions at 14~TeV will be recorded with up to $\mu=200$ proton-proton interactions per bunch crossing. 

In order to cope with  a threefold increase of $\mu$  with respect to Run 2 and to enable efficient pileup mitigation, all detector subsystems must be upgraded or replaced. 
The planned upgrade is described in the Letter of Intent~\cite{ATLAS:1502664}, where various layout options are considered, 
and summarised in the Scoping Document~\cite{CERN-LHCC-2015-020} (2015), where the final layout was chosen based on desirable physics capabilities.
%Increased $\mu$ forces updates in the trigger. 
In the trigger and data acquisition system a two-level hardware trigger with L0 output rate up to 1~MHz and L1 up to 400~kHz is foreseen together with
a High-Level Trigger with 10 kHz output (permanently recorded data).
``Custom hardware'' triggers will enable data streaming at rates 1-40 MHz. 
%New L0/L1 triggers in the Inner Tracker, Calorimeters and Muon Spectrometer will be installed.
%Inner tracker has to cope with increased radiation and
The inner tracker itself will be replaced with a new, all-silicon  detector.
Its  pseudorapidity coverage will be increased from the current  $|\eta| \leq 2.4$ to $|\eta| \leq 4$
 by the addition of four pixel discs in the forward region.
Since calorimeter cell noise (electronic and pileup) increases with pseudorapidity $|\eta|$, detector improvements in the forward regions are necessary. 
The Liquid Argon (LAr) forward electromagnetic calorimeter with be replaced by a  higher granularity detector (sFcal).
Additionally, a High Granularity Timing Detector will be installed in front of LAr end-caps, in the  pseudorapidity region 
2.4 $\leq |\eta| \leq$ 4.3. Readout electronics of LAr and Tile Calorimeters will be replaced due to forseen radiation damage in Run 3.
For the muon system addition of new Resistive Plate Chambers (RPCs) in the barrel is expected.

In these proceedings I will compare physics prospects for the HL-LHC with 
prospects for 300 fb$^{-1}$ integrated luminosity expected at Run 3. 
Section 2 will decribe the procedure of extracting the SM Higgs coupling strengths mentioning assumptions behind different fits. I will point out the role of theoretical uncertainties in improving the precision on coupling extraction. 
In section 3 searches for signs of BSM physics in Higgs couplings will be outlined.
In section 4 I will discuss prospects for observing Higgs pair production and measuring Higgs trilinear self-coupling.
An example of a search for supersymmetric particles will be discussed in Section 5.

The physics studies described in these proceedings have been performed in the period 2014-2015, hence not all improvements of the upgraded detector 
were taken into account in every analysis. Furthermore, some analyses assume earlier estimates on instantenuous luminosity and $\mu=140$.

%%%%%%%%%%%%%%%%%%%%%%%
\section{Measurement of SM Higgs couplings}

\begin{figure}
\center
\includegraphics[width=6cm]{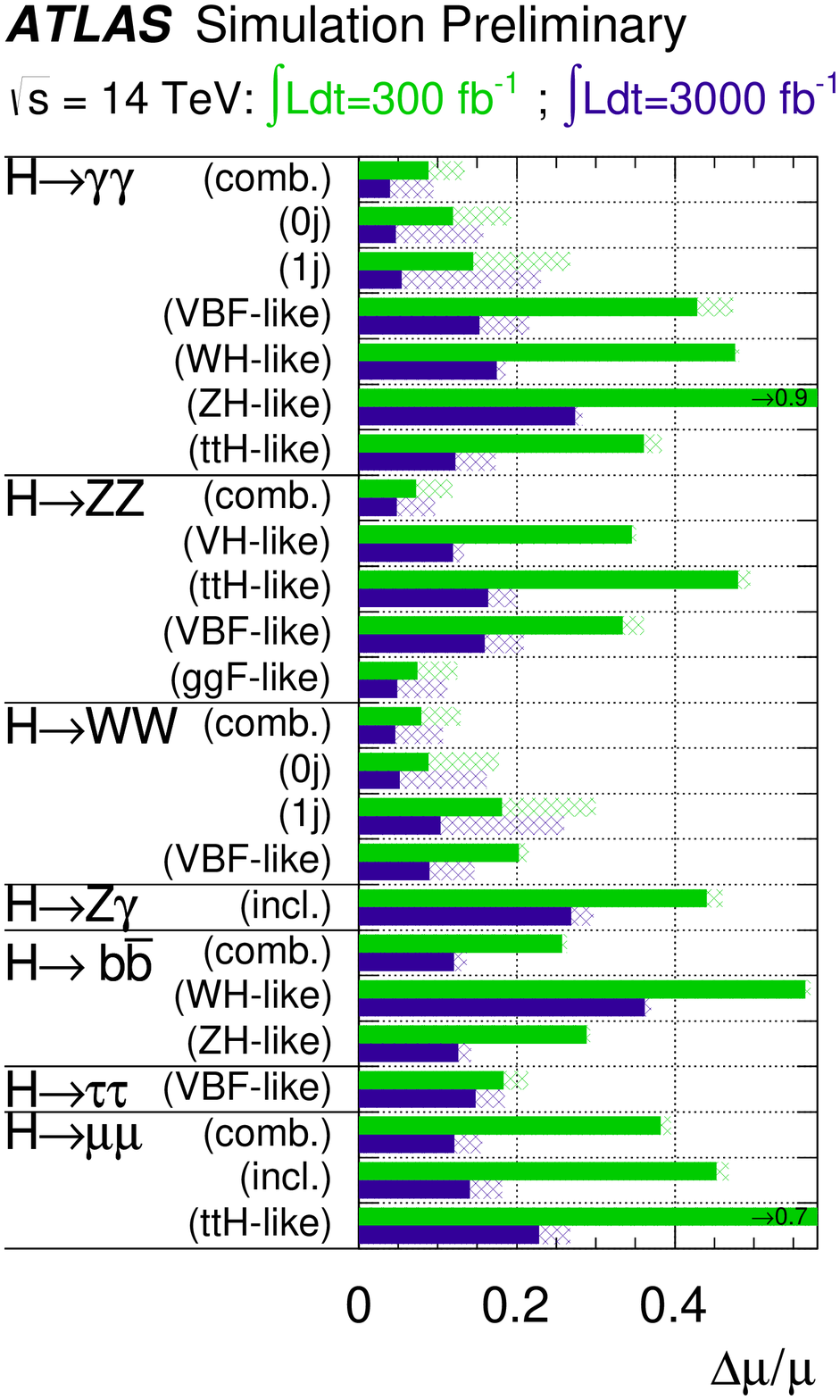}
\includegraphics[width=6cm]{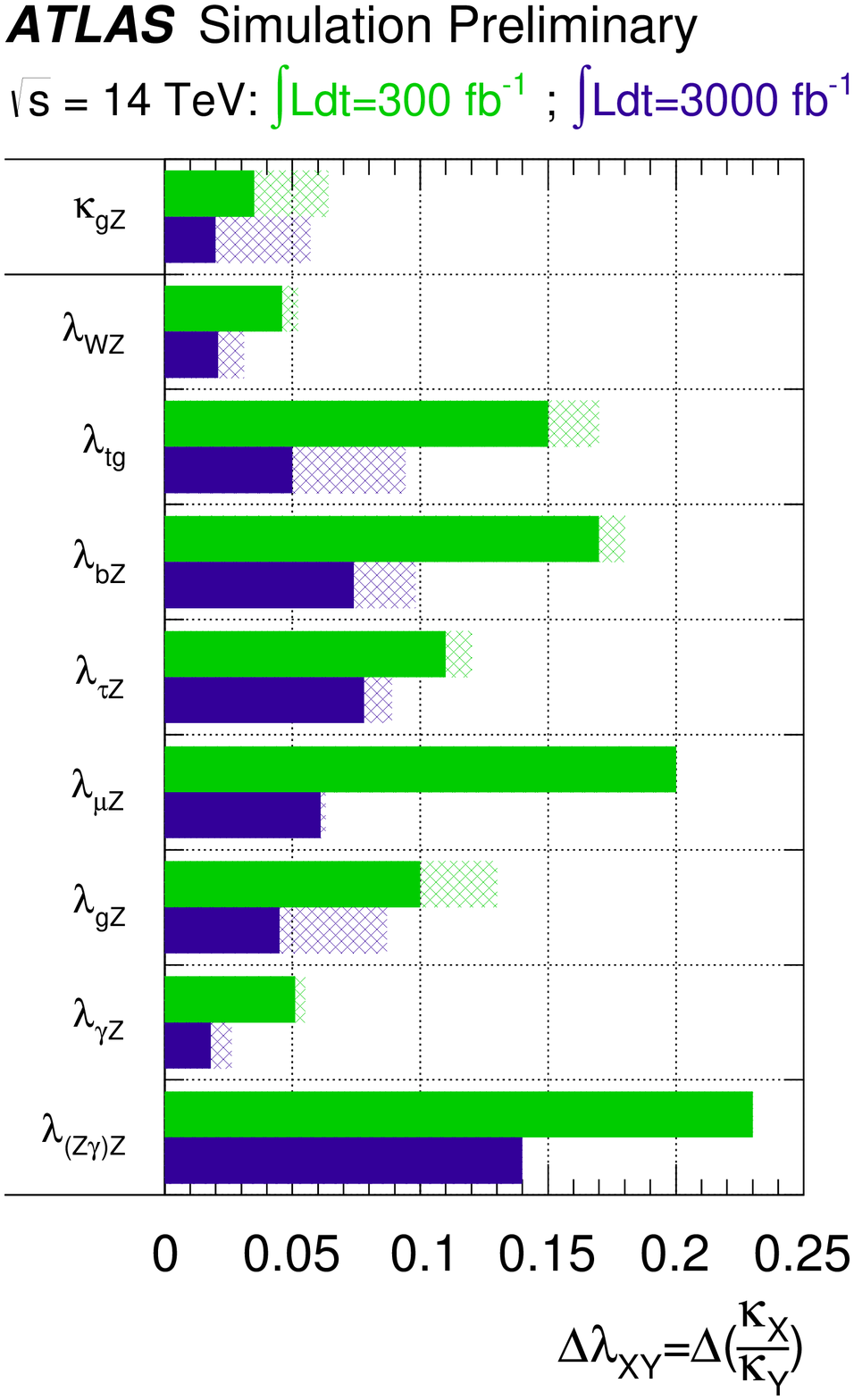}
\caption{The uncertainties of the Higgs signal strengths in different categories (left)
 and couplings ratios $\lambda_{ij}$ (right). Figures taken from~\cite{ATL-PHYS-PUB-2014-016}.}
\label{fig:cstrenghts}
\end{figure}

The HL-LHC provides an opportunity to test the Standard Model with precise measurements of the Higgs boson properties.
ATLAS analysed the following channels: 
$h \to \mu \mu$, $h\to \tau \tau$, $h \to ZZ$, $h \to WW$, $h \to \gamma \gamma$ (0/1/2 jet categories, inclusive), 
$h \to Z \gamma$, $Vh, t \bar{t} h \to \gamma \gamma $, $Vh \to b \bar{b}$~\cite{ATL-PHYS-PUB-2014-016}.
%The precision of signal strength $\mu$ is given as relative uncertainty $\Delta\mu/\mu$.
The precision for each of the Higgs boson production and decay  categories ($\Delta\mu/\mu$) is presented 
in the left panel in Fig.~\ref{fig:cstrenghts}.  Green bars correspond to 300~fb$^{-1}$ and dark blue bars correspond
to 3000~fb$^{-1}$. The highest precision of $\sim$ 5\% can be reached with 3000~fb$^{-1}$ in di-boson final states. 
An experimental precision of $\sim$ 4\% is reachable for the dominanant producion mode, the gluon fusion.
The largest gains in precision between 300 and 3000~fb$^{-1}$ is observed in rare channels: 
production associated with vector bosons and associated with $t\bar{t}$ because these channels are  statistically dominated.

%With the forseen luminosity even rare Higgs producion and decay channels become observable.
%couplings to bosons and fermions. Together with the mass measurement they 
The measurements of signal strenghts are interpreted as the SM LO Higgs couplings to bosons and fermions. 
%kappa framework
Assuming the decay width of the Higgs boson to be zero, the coupling strengths in its production and decay can be factorised. 
Deviations from the SM Higgs production cross-sections $\sigma $ and its branching ratios $B$ are denoted
as $\kappa_i$  and $\kappa_j$ in the following way:
\begin{equation}\label{eq:ratios}
\frac{\sigma(ii\to h) B(h\to jj)}{\sigma_{SM}(ii\to h) B(h\to jj)_{SM} }
=
\frac{\kappa_i ^2 \kappa_j^2}{\kappa_h^2},
\end{equation}
where $\kappa_h$ denotes the scaling of the SM Higgs width $\Gamma_h$.
All $\kappa_i$ parameters correspond to LO degrees of freedom.

To avoid large uncertainties related to $\Gamma_h$, assumptions on Higgs branching ratios to invisible particles  and any other model-dependent theoretical and experimental assumptions in the coupling fit
some parametrisations use ratios $\lambda_{ij} = \frac{\kappa_i}{\kappa_j}$.
%\begin{equation}\label{eq:ratios}
%\lambda_{ij} = \frac{\kappa_i}{\kappa_j}.
%\end{equation}
The expected precision on these ratios are shown in the right panel in Fig.~\ref{fig:cstrenghts}.

In the minimal coupling fit it is assumed that all Higgs  couplings to bosons, $\kappa_V$, 
and fermions, $\kappa_f$, are identical.
In this model the expected experimental precision with 300 fb$^{-1}$  can reach $\sim$2.5\% on  $\kappa_V$ 
and $\sim$7.1 \% on $\kappa_f$. 
Including current theory uncertainties the total uncertainties of $\sim$4.3\% and $\sim$8.8\%
are expected. The largest reduction of uncertainties with the increased luminosity is expected for $\kappa_f$,
where an experimental precision of $\sim$3.2\% can be reached ($\sim$5.1\% including current theory uncertainties)
compared to $\kappa_V$ $\sim$1.7\% ($\sim$3.3\% including theory uncertainties).
%%%%%%%%%%%%%%%%%%%%%%%
\section{BSM Physics in Higgs couplings}

With the estimated precision of measuring Higgs couplings at 300 and 3000 fb$^{-1}$  
constraints on theories that extend the Standard Model
 are investigated~\cite{ATL-PHYS-PUB-2014-017}: 
an additional electroweak singlet, an additional electroweak doublet (i.e. the two-Higgs-doublet model), a composite Higgs boson, a simplified Supersymmetric Standard Model (MSSM), and a Higgs portal to dark matter.
I will highlight one of these results here:  tests of models predicting Higgs compositeness.
 
If the Higgs is not an elementary particle but a composite pseudo-Goldstone boson 
its couplings to bosons and fermions will deviate from the SM predictions. 
These deviations can be parametrised in terms of a Higgs compositeness scale $F$ and Higgs vacuum expectation value $v$: 
$\xi = v^2/F^2$. In the SM $\xi$ is zero as $F \to \infty$.
Simple models predicting a composite  Higgs boson include MCHM4 and MCHM5~\cite{MCHM}.
In MCHM4  the ratios of the Higgs boson couplings to bosons and fermions to their SM expectations
are equal and can be parametrised by
\begin{equation}
\kappa_V = \kappa_f = \sqrt{1-\xi}.
\end{equation}
MCHM5 predicts  
\begin{equation}
 \kappa_{V} =  \sqrt{1-\xi}, \quad \quad \kappa_f =\frac{1-2\xi}{\sqrt{1-\xi}}
\end{equation}

A comparison of exclusion powers for these models, assuming the SM Higgs boson,  for the Run 3 and HL-LHC 
data is presented in 
%Figure~\ref{fig:MCHM_CL}
Table \ref{tab:compositenessscale}.
%\begin{figure}
%\center
%\includegraphics[width=6cm]{MCHM_couplingsPlot.png}
%\caption{Expected 95\% CL lower limits on Higgs compositeness scale.}\label{fig:MCHM_CL}
%\end{figure}
%Theoretical uncertainties play a large role in the exclusion power. Table~\ref{tab:compositenessscale}

\begin{table}
\center
\begin{tabular}{|c r r | r r |}
%\begin{tabular}{| c c c |}
\hline
Model & &300 fb$^{-1}$ & &3000 fb$^{-1}$ \\
%\end{tabular} \\
%\begin{tabular}{|c r r | r r |}
& All unc. & No theor. unc. & All unc. & No theor. unc.\\
\hline
MCHM4 & 620   & 810  & 710   & 980 \\
MCHM5 & 780   & 950  & 1000 &1200\\
\hline
\end{tabular}
\caption{Expected 95\% CL lower limits on Higgs compositeness scale $F$ in GeV.
Table taken from~\cite{ATL-PHYS-PUB-2014-017}.}\label{tab:compositenessscale}
\end{table}
%%%%%%%%%%%%%%%%%%%%%%%
\section{Higgs Pair Production}

A direct insight into the mechanism of electroweak symmetry breaking could be gained by  the measurement of Higgs potential $V$.
In the Standard Model after electroweak symmetry breaking %the Standard Model Higgs potential 
\begin{equation}
V=\frac{m_h^2}{2}h^2 + \lambda_{hhh} h^3 + \frac{\lambda_{hhhh}^4}{4}h^4.
\end{equation}
The Higgs boson acquires triple and quartic self-couplings,  fully determined by the Higgs mass
$m_h$ and vacuum expectation value $v$: 
$\lambda_{hhh}=m_h^2/(2v)$ and $\lambda_{hhhh}=m_h^2/(2v^2)$.
%The measurement of the Higgs self-couplings will enable to determine if
% is indeed triggered by the SM Higgs boson.
Any deviations from these values will indicate that 
electroweak symmetry breaking is triggered by a more complex physical mechanism than that of Ref.~\cite{Higgs mechanism}.

The direct measurement of Higgs trilinear (quartic) coupling involves the observation of two (three) Higgs bosons in the final state.
The cross-section of di-Higgs  production in the dominant gluon-gluon fusion  production mode		
is small $\sigma^{NNLO+NNLL}=39.56$ fb~\cite{hh} The tri-Higgs production cross-section yields only 
 $\sigma^{NLO FTapprox} = 0.09$ fb~\cite{hhh}. Therefore di-Higgs  production will probably be observed for the first time at the HL-LHC, while tri-Higgs production is beyond its reach.
%The observation of triple Higgs production will be challenging even with 3000 fb$^{-1}$.
 ATLAS has performed two searches for the SM Higgs pair production: in the $b\bar{b}\gamma\gamma$~\cite{ATL-PHYS-PUB-2014-019}, 
 $b\bar{b}\tau \tau$~\cite{ATL-PHYS-PUB-2015-046} channels. 

%The $b\bar{b}b\bar{b}$ chanels benefits from  the largest branching fraction. The dominant backgrounds 
%originate from multi-jet processes.
The $b\bar{b}\gamma\gamma$ channel benefits from good mass resolution of the Higgs decaying to two photons and from relatively small backgrounds. The challenge of this analysis lies in small signal yields: at the HL-LHC about 300 events are expected. In addition to  reducible backgrounds --
continuum $b\bar{b}\gamma\gamma$, single Higgs: $t\bar{t}H(\to \gamma\gamma)$,
$Z(\to b\bar{b})H(\to\gamma\gamma)$ and  $b\bar{b}H(\to\gamma\gamma)$ --
the signal is accompanied by reducible backgrounds from light-flavour jets misidentified as $b$-jets and 
electrons misidentified as photons.
This analysis assumes $\mu= 140$ and estimates the upgraded ATLAS performance for
measuring the $\eta$ and $p_T$ of jets and photons with the 
use of the $\eta/p_T$-dependent eficiency and resolution functions. 
The transverse momenta of the jets are smeared by 10-25\%.
The performance of b-tagging for these jets is simulated by applying an efficiency function, 
 corresponding to a mean efficiency of 70\%.
 To to account for radiation of partons outside the jet cone and semi-leptonic $b$-hadron decays a correction of jet energy is applied. 
The identification of photons is modelled with a plateau efficiency equal to 76\% for photons with transverse momenta larger than 80 GeV.
The rates of jets being misreconstructed as photons, light-flavour jets as b-jets and 
c-jets as b-jets are assumed to be equal to 0.25\%, 1\%, and 30\%, respectively.
In the event selection two energetic, isolated photons and two b-jets are required. 
The invariant mass of two photons (two b-jets) must fulfill the requirement  123 GeV $< m_{\gamma\gamma}<$ 128 GeV (100 GeV$<m_{b\bar{b}}$ < 150 GeV) and transverse momenta
$p_T^{\gamma\gamma}, p_T^{b\bar{b}} $ > 110 GeV.
The effect of systematic uncertainties is neglected in this analysis.
The expected limits set on the value of the Higgs trilinear coupling $\lambda$ (assuming no other sources of new physics) 
are $\lambda/\lambda_{SM} \in [-1.3, 8,7]$ at 95\% C.L..
%$\lambda/\lambda_{SM}\leq -1.3$ and $\lambda/\lambda_{SM}\geq 8.7$.

The $b\bar{b}\tau \tau$ channel has the third largest di-Higgs branching ratio of $\sim$7.3\%. The search 
in this channel is performed in three subchannels defined
by decay modes of the $\tau$ leptons: di-leptonic, semi-leptonic and hadronic.
This analysis assumes $\mu=200$ and uses a track a confirmation algorithm to suppress pile-up.
The dominant backgrounds are the production of $t\bar{t}$, $W$ + jet, $Z$+jet,  di-bosons and multijet events.
The largest sensitivity is obtained in the hadronic subchannel, resulting in an upper
limit on the Higgs pair production cross-section equal to 4.3 times its SM value at 95\% C.L.
The corresponding expected exclusion limits on $\lambda_{hhh}$ in all three subchannels are 
$\lambda/\lambda_{SM}\in[-4, 12]$.
%$\lambda/\lambda_{SM}\leq -4$ and $\lambda/\lambda_{SM}\geq 12$.

%depicted in 
%Fig.~\ref{fig:lambdahhh}.
%\begin{figure}
%\center
%\includegraphics[width=6cm]{fig_08}
%\includegraphics[width=6cm]{fig_09}
%%\includegraphics[width=6cm]{hh_bbyy_limitslambda_fig_08.png}
%%\includegraphics[width=6cm]{hh_bbtautau_limitslambda_fig_09.png}
%\caption{Limits on the $\lambda_{hhh}$ projected at the HL-LHC in $b\bar{b}\gamma\gamma$ and $b\bar{b}\tau \tau$ channels.}\label{fig:lambdahhh}
%\end{figure}

Higgs pair production can also probe the presence of TeV scale resonances interacting with  the Higgs, such as Kaluza-Klein gravitons. Such a  resonant production of a pair of Higgs bosons  decaying to $b\bar{b}b\bar{b}$ was studied in~\cite{CERN-LHCC-2015-020}. 
The HL-LHC study found a 4.4$\sigma$ expected significance for the graviton with a mass of  2 TeV.
The largest gain in sensitivity between different inner tracker layouts considered in the Scoping Document
stems from an inproved  $b$-tagging performance.

%%%%%%%%%%%%%%%%%%%%%%%
\section{Prospects for discovering supersymmetric electro-weakinos}

A  possible explanation of  the electroweak symmetry breaking explored by ATLAS
is through a supersymmetric extension of the Standard Model.
This theory predicts the existence of many new particles, among them charginos 
$\tilde{\chi}^{\pm}_1$ and neutralinos $\tilde{\chi}^{0}_2$, the mass eigenstates from superposition 
of the supersymmetric  Higgs partners and electroweak gauge bosons. 
The ATLAS sensitivity to detecting electro-weakinos in the mass range of several hundreds GeV 
was studied in a simplified Wh-mediated model~\cite{ATL-PHYS-PUB-2014-010, ATL-PHYS-PUB-2015-032}, in which 
$\tilde{\chi}^{\pm}_1$ and $\tilde{\chi}^{0}_2$ are pair-produced  and decay to the W boson and the lightest neutralino $\tilde{\chi}^{0}_1$,
 Higgs boson and $\tilde{\chi}^{0}_1$, respectively. 
The studied  experimental signature consists of a lepton, a pair of b-quarks and missing energy due to undetected neutrino and neutralinos.

In this analysis $\mu=$60 ($\mu=$140) was assumed for Run 3 (HL-LHC).
With the 300fb$^{-1}$ (3000 fb$^{-1}$) ATLAS is sensitive to electro-weakinos down to 1.1 TeV (800 GeV) mass range. The improvement in the HL-LHC phase is driven by increase in b-tagging efficiency due to upgrades in the inner tracker and calorimeter.

%%%%%%%%%%%%%%%%%%%%%%%
\section{Conclusions}

Understanding electroweak symmetry breaking requires precise measurements in the Higgs sector.
In order to accomplish this task in the challenging environment of the HL-LHC
extensive studies of physics prospects are carried out to optimise the planned detector upgrades.

Higgs couplings to bosons and fermions can be measured at HL-LHC
with  uncertainties up to two times smaller  than with 300 fb$^{-1}$. The largest improvements are expected in rare VH and ttH production channels.  The precision of measurements is limited by theoretical uncertainties of the gluon fusion production cross-section. 
These measurements are also interpreted in the context of BSM models. 
In the MCHM5 discussed in these proceedings the lower limit on the Higgs compositeness scale can be set at 1 TeV with 3000 fb$^{-1}$, compared to 780 GeV with 300 fb$^{-1}$. 

The direct probe of the Higgs potential --  di-Higgs production --  could be observed for the first time at the HL-LHC even in the rare $b\bar{b}\gamma\gamma$ channel. 
In the final states with $b\bar{b}\gamma\gamma$ and $b\bar{b}\tau\tau$ 
  the $\mathcal{O}(1)$ limits can be set on Higgs trilinear coupling $\lambda_{hhh}$.

The discovery potential  of direct searches for New Physics interacting with the Higgs will be extended. 
 Sensitivity to Kaluza-Klein Graviton reaches up to 2 TeV.
ATLAS is sensitivie to electro-weakinos in the 800 GeV mass range.
For all analyses theory improvements are crucial in obtaining higher precision.

\end{document}